\documentclass[preprint,5p,times,twocolumn ]{elsarticle}

\usepackage{amssymb}
\usepackage{amsmath,amssymb}
\usepackage{type1cm}
\usepackage{mathrsfs}
\usepackage{url}
\usepackage{graphicx}
\usepackage{color}
\usepackage{comment}

\newcommand\HL[1]{{\color{black}#1}}

\usepackage[columnwise]{lineno}

\def\vec#1{\mbox{\boldmath $#1$}}
\journal{JPCO}
\begin{document}

\begin{frontmatter}



\title {
Throughput reduction on an air-ground transport system by the simultaneous effect of multiple traveling routes equipped with parking sites
}



\author[RCAST]{Satori Tsuzuki}
\ead{tsuzuki.satori@mail.u-tokyo.ac.jp}

\author[RCAST]{Daichi Yanagisawa}
\ead{tDaichi@mail.ecc.u-tokyo.ac.jp}

\author[RCAST]{Katsuhiro Nishinari}
\ead{tknishi@mail.ecc.u-tokyo.ac.jp}

\address[RCAST]{Research Center for Advanced Science and Technology, The University of Tokyo, 4-6-1, Komaba, Meguro-ku, Tokyo 153-8904, Japan}

\begin{abstract}
This paper examines the traffic flows on a two-dimensional stochastic lattice model that comprises a junction of two traveling routes: the domestic route and the international route each of which has parking sites.
In our model, the system distributes the arrived particles to either of the two routes and selects one of the parking sites in the route for each particle, which stops at the parking site once during its travel. Because each particle has antennas in the back and front directions to detect other approaching particles, the effect of the volume exclusion of each particle extends in the moving direction. The system displays interesting behavior; remarkably, the dependence of the throughput on the distribution ratio of particles to the domestic route reduces after reaching the maximum parking capacity of the domestic route. Our simulations and analysis with the queueing model describe this phenomenon and suggest the following fact: \HL{As the distribution ratio of particles to the international route decreases, the throughput of the international route reduces, and simultaneously, that of the domestic route saturates. The simultaneous effect of the decrease and saturation causes a reduction in the throughput of the entire system.}
\end{abstract}


\end{frontmatter}





\section{INTRODUCTION}

The stochastic lattice model, which was started with the cellular automata (CA) by Neumann and Ulam~\cite{Neumann:1966:TSA:1102024}, has been acknowledged in studies on traffic flow problems.  Above all, the totally asymmetric simple exclusion process (TASEP) has successfully described various kinds of traffic flow systems ranging from molecular mechanics~\cite{Teimouri2015, Denisov2015} to vehicular traffic systems~\cite{PhysRevE.89.042813, 1742-5468-2017-4-043204, 1751-8121-42-44-445002, 2010arXiv1001.4124Y, doi:10.1142/S0218202515400011}, all because of a simple mechanism: each particle hops to the neighboring cite in a one-way direction only when the site is empty. 

Until today, several studies of traffic flow at the junction on stochastic lattice models have been reported, as exemplified by~\cite{1674-1056-19-9-090202, 0295-5075-80-6-60002, doi:10.1142/S0217984911027339} in the fundamental studies and~\cite{CDA14, 6248222} in application fields. Most of them focus on systems that connect the single lane that has no adjacent lane. In real-world cases, however, many systems comprise a junction of multiple lanes, each of which has parking sites (e.g., parking areas on highways, and aircraft taxiing at the airport). Another research groups studied the traffic flows on parallel multiple lane systems~\cite{Ezaki2011PositiveCE, Ichiki2016,Verma2015, PhysRevE.70.046101, doi:10.7566/JPSJ.85.044001, Yanagisawa2016, 0295-5075-107-2-20007, RePEc:wsi:ijmpcx:v:18:y:2007:i:09:p:1483-1496}. In contrast, these studies do not focus on junction flows.

The goal of this study was to explore the mechanisms of traffic flows on a two-dimensional stochastic lattice model that comprises a junction of two traveling routes, each of which has adjacent parking sites. The whole system distributes arrived particles to either of the traveling routes and selects one of the parking sites of the route for each particle. The particle stops at the designated site of the route once during its travel. When a particle finds that the targeted site is in use, the particle changes its destination to one of the other parking sites that is vacant.

As preliminary works, we investigated the characteristics of the one-dimensional stochastic lattice model that has a single traveling lane equipped with the functions of site assignments to the adjacent parking sites~\cite{PhysRevE.97.042117, PhysRevE.98.042102}. Our current model can be said to be a combined model of our two previous models in a different scale in a broader sense; however, the existence of a single junction and the extended volume exclusion effect are the distinguishing factors compared to the previous models. 

The remainder of this paper is structured as follows. Section 2 describes our model with the priority rules at each intersection and an effective method to detect the approaching particles. In Section 3, we investigate the characteristics of the system properties through simulations and compare them with the theoretical queueing models~\cite{ERLANG-A-K}. Section 4 summarizes our results and concludes this paper.

\section{MODEL}
\subsection{Basic concept of our model}
\begin{figure}[t]
\begin{center}
\vspace{-5.0cm}
\includegraphics[width=\textwidth, clip, bb= 0 0 2026 1127]{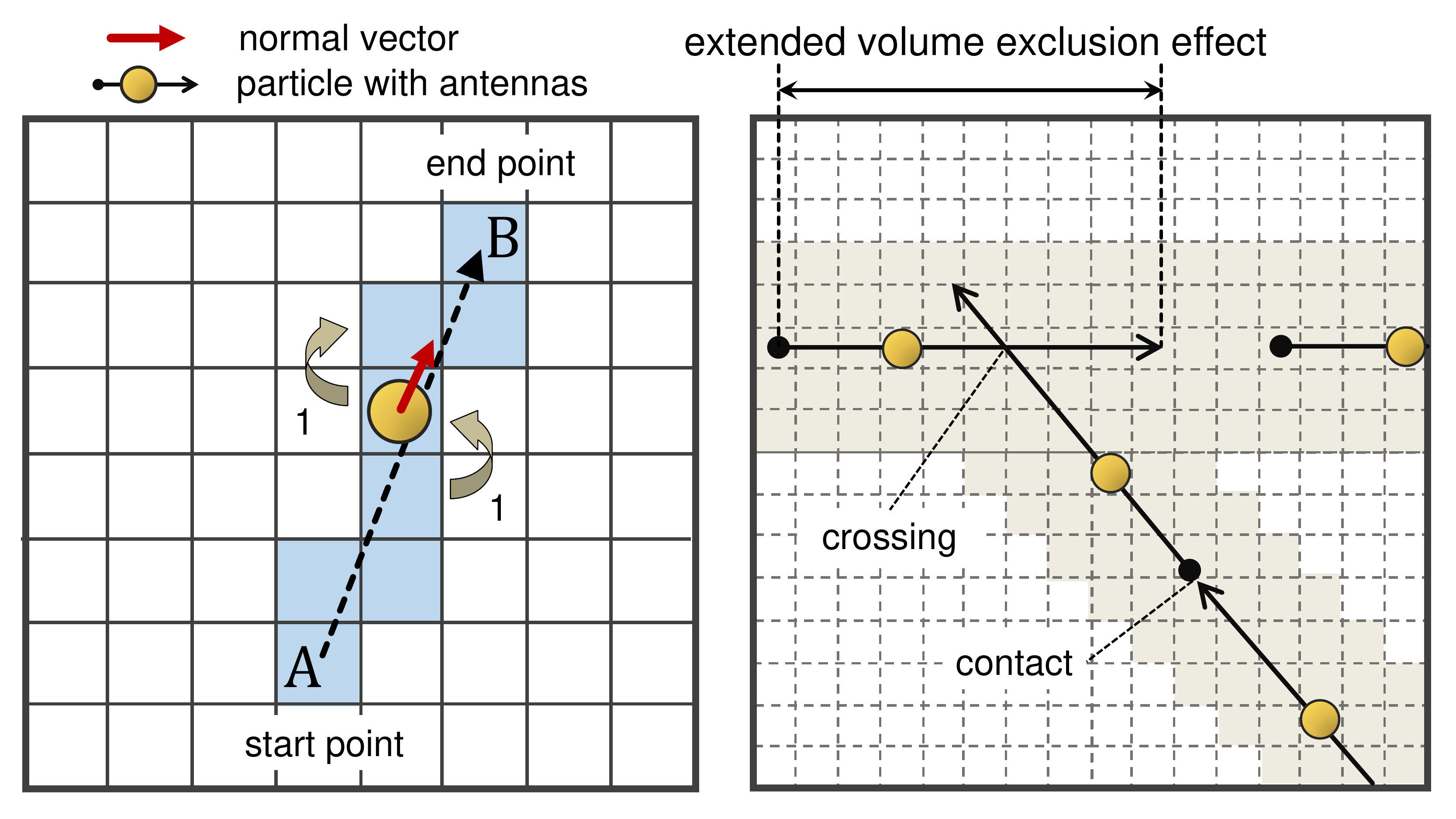}
\end{center}
\caption{Basic concept of our model.}
\label{fig:schemview1}
\end{figure}

Figure \ref{fig:schemview1} depicts the basic concept of our model. A checkerboard spread over the whole system, and the selected cells from the checkerboard, called checkpoints, construct a route. In the example of the left part in Fig.\ref{fig:schemview1}, cells A and B represent the start and end cells, respectively. The particle hops from cell A to cell B along the relative vector obtained by connecting the centers of both cells. At this time, the particle moves along the vector by a distance equal to the length of a single cell in every time step and stores itself on the closest cell from the relative vector.

In our model, each particle has a pair of antennas in the back and front directions as shown in the right part of Fig.\ref{fig:schemview1}. We detect the intersections and the contacts of two different antenna particles by using the polyhedral geometric algorithm as mentioned later in Section~\ref{sec:detectalg}. When the system detects the intersection, the system instructs both or either of the particles to stop at the current location, according to the priority rules predefined at the junction.

\subsection{Target system}
\begin{figure*}[htb]
\vspace{-11.0cm}
\hspace{+1.5cm}
\includegraphics[width=1.7\textwidth, clip, bb= 0 0 2026 1127]{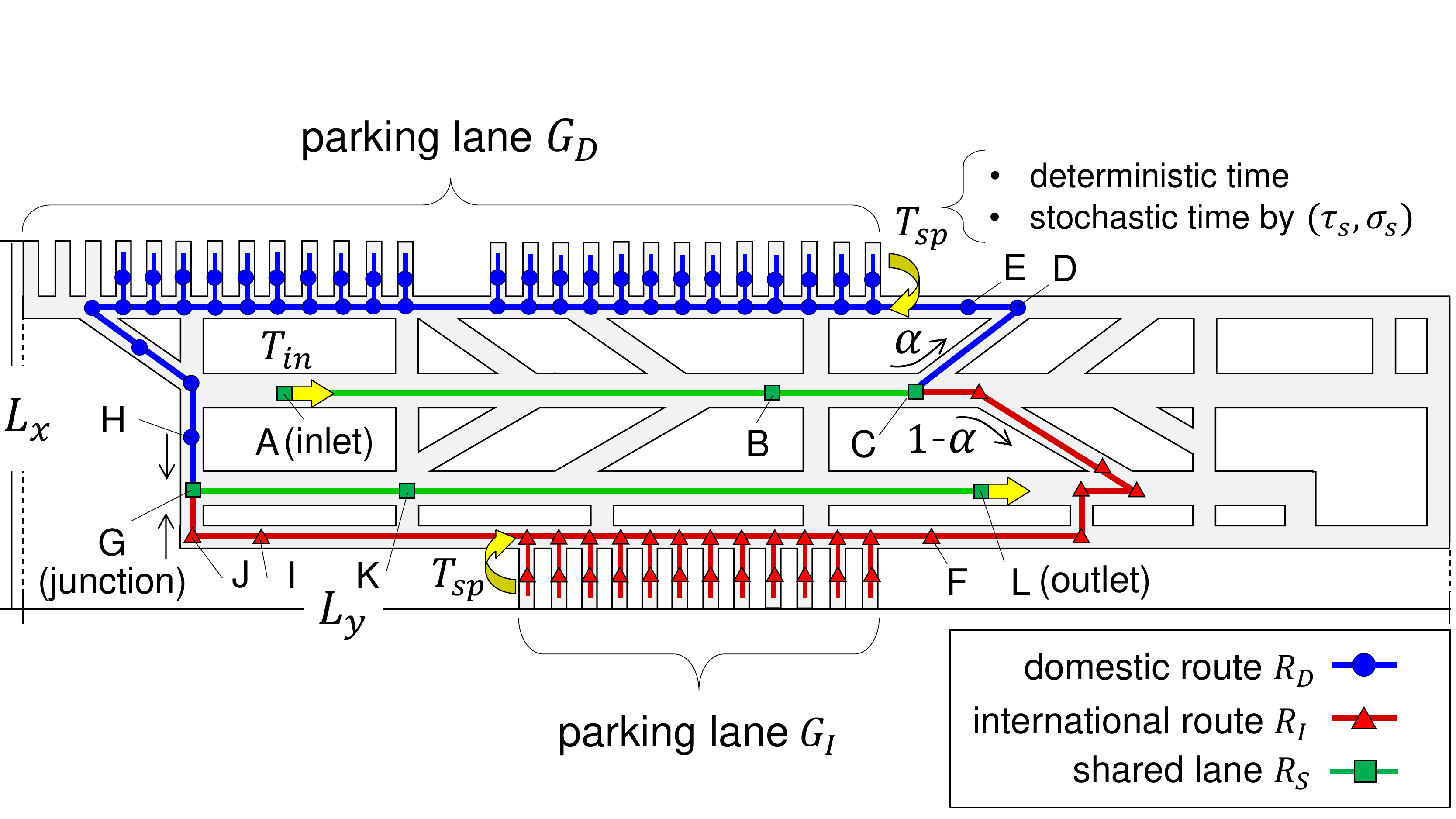}
\caption{Schematic of the target system.}
\label{fig:schemview2}
\end{figure*}

Figure~\ref{fig:schemview2} depicts the schematic of the target system.
We abstract the target system from a real-world airport, Fukuoka airport in Japan, which has one of each domestic and international terminals; 
the roads on the background show the coarse-grained geometries of the airport for reference.
We denote the number of cells in the x-direction by $L_{\rm x}$ and denote those in the y-direction by $L_{\rm y}$.
The three symbols~(circle, square, and triangle)~indicate the checkpoints on the checkerbord. 
The blue line with the circular symbol shows a domestic route $R_{\rm D}$ comprising the parking lane $G_{\rm D}$ that has $23$ parking sites, and 
the red line with the triangle symbol shows a international route $R_{\rm I}$ comprising the parking lane $G_{\rm I}$ that has 12 parking sites. 
The upper and lower green lines with square symbols respectively show the lanes utilized for the arrival and departure of particles.  

Before the simulation, we establish a timetable that has five instructions for each particle: (a) the arrival time $t_{\rm arr}$~determined by the fixed interval of arrival time $T_{\rm in}$, (b) the type of routes (the domestic route $R_{\rm D}$~or~the international route $R_{\rm I}$), (c) the index of the parking site of the route, (d) the scale and shape parameters that determine the staying time $T_{\rm sp}$ as described later, and (e) the interval velocities among the checkpoints. 
We set the arrival time $t_{\rm arr}$ of each particle such that all intervals of the arrival times become the same constant value $T_{\rm in}$.
In additions, we assign the domestic route $R_{\rm D}$ to particles with probability $\alpha$, and we assign the international route $R_{\rm I}$ to them with the probability $1-\alpha$. 
In the end, we select one of the parking sites on the route for each particle at uniform random.

During the simulation, each particle enters the system from inlet A with a fixed interval $T_{\rm in}$ and moves on the path ABC towards the checkpoint C. 
The system distributes the particle to either of the routes according to the timetable at checkpoint C. 
At checkpoints E or F, the particle checks the state of the parking site designated in the timetable and changes it
when the parking site is still in use. In such a case, the particle selects one of the parking sites on the route from the rest of the parking sites whose state is vacant. Then, the particle moves towards the parking site.

After reaching the parking site, the particle stays at the parking site during time $T_{\rm sp}$. 
In this paper, we investigate the target system in both cases with stochastic and deterministic parameters to determine the effect of the delay of staying time $T_{\rm sp}$.
In the former case, we consider the delay of the staying time $T_{\rm sp}$ by using a half-normal distribution. 
The half-normal distribution is selected because the delay only occurs for the positive direction in the target system. 
The scale and shape parameters ($\tau_{\rm s}$, $\sigma_{\rm s}$) of the half-normal distribution are obtained from the mean and deviation of ($\bar{\tau}_{\rm s}$, $\bar{\sigma}_{\rm s}$) of a normal distribution, as follows: 
\begin{eqnarray}
\tau_{\rm s}   &=&  \bar{\tau}_{\rm s} + \bar{\sigma}_{\rm s} \sqrt{\frac{2}{\pi}}\label{eq:half-ndist3}\\ 
\sigma_{\rm s} &=& \bar{\sigma}_{\rm s} \sqrt{(1-\frac{2}{\pi})} \label{eq:half-ndist4} 
\end{eqnarray}
In the latter case, we decide the $T_{\rm sp}$ by setting $\bar{\sigma}_{\rm s}$ to zero.

The particles move toward checkpoint G after vacating the parking sites. The flows of the particles from both, the domestic route $R_{\rm D}$ and international route $R_{\rm I}$, merges at checkpoint G. When some particles go through path HG on the domestic route $R_{\rm D}$, other particles move on path IJG, and the system pauses the particles on path HG. After passing checkpoint G,  the particle moves on path GKL and exits from the outlet L. Note that we describe the setting of interval velocities among the checkpoints in Section.\ref{seq:inputpara}
.

\subsection{Detection of approaching antenna particles}\label{sec:detectalg}
In this section, we describe a polyhedral geometrics algorithm for the detection of approaching antenna particles.
Let the number of particles that exist on the system be $N_{\rm p}$. We set a sequential number from $1$ to $N_{\rm p}$ to the particles.
We detect the collision of antennas among the $i$-th particle and the other $j$-th particle, as follows. 

First, we judge whether the $i$-th and $j$-th particles are on an identical line by using the relative vector from the $i$-th particle to the $j$-th particle ${\vec{x}}_{\rm ij}$, 
   the normal vector of $i$-th particle ${\vec{n}}_{\rm i}$, 
   \HL{the angle $\phi = \Bigl |\arccos \Bigl\{\frac{(\vec{n}_{\rm i}\cdot \vec{x}_{\rm ij})}{|\vec{n}_{\rm i}||\vec{x}_{\rm ij}|} \Bigr \} \Bigr |$~
   (the angle between the normal vector $\vec{n}_{\rm i}$ and the relative vector $\vec{x}_{\rm ij}$), 
   and the small angle $\epsilon$. 
   If the condition:}
\begin{eqnarray}
\phi~~<~~\epsilon~~\rm{or}~~\phi~~>~~\pi-\epsilon \label{eq:onsameroad}
\end{eqnarray}
\HL{is satisfied, we judge that the $i$-th and $j$-th particles are on an identical line.}
$\epsilon$ becomes zero in an ideal case; however, we must set it to a larger value than the scale of the ``machine epsilon.'' 

In case $\phi$ meets the relationship in Eq.(\ref{eq:onsameroad}), 
   we decide that the $j$-th particle is on an identical line as the $i$-th particle, which indicates that these two particles are on the same path.
In this case, we can judge the collision of the antennas by simply measuring the distance between both particles.
At this time, we have to identify the back and front positions of the two particles and their orientations because the system needs to issue a pause instruction to the back particle. By judging from the normal vector of the $i$-th particle $\vec{n}_{\rm i}$, of the $j$-th particle $\vec{n}_{\rm j}$, and the relative vector $\vec{x}_{\rm ij}$, four interaction types exist, as listed below.

\begin{description}
\item [(A)]~If $(\vec{n}_{\rm i} \cdot \vec{n}_{\rm j}) > 0$ and $(\vec{n}_{\rm i} \cdot \vec{x}_{\rm ij}) > 0$, then:\\ The $i$-th particle locates the back of the $j$-th particle. The system issues a pause instruction to the $i$-th particle.

\item [(B)]~If $(\vec{n}_{\rm i} \cdot \vec{n}_{\rm j}) > 0$ and $(\vec{n}_{\rm i} \cdot \vec{x}_{\rm ij}) < 0$, then:\\ The $j$-th particle locates the back of the $i$-th particle. The system issues a pause instruction to the $j$-th particle.

\item [(C)]~If $(\vec{n}_{\rm i} \cdot \vec{n}_{\rm j}) < 0$ and $(\vec{n}_{\rm i} \cdot \vec{x}_{\rm ij}) > 0$, then:\\ Both particles face each other. 
This case only occurs at junction G in the target system. The system issues a pause instruction to the either of the particles according to the priority rule predefined at junction G. 

\item [(D)]~If $(\vec{n}_{\rm i} \cdot \vec{n}_{\rm j}) < 0$ and $(\vec{n}_{\rm i} \cdot \vec{x}_{\rm ij}) < 0$, then:\\ Both particles move in the opposite direction. Since this situation does not occur in the target system, the system ignores this type of interaction.
\end{description}
\begin{figure*}[t]
\vspace{-9.0cm}
\hspace{+2.0cm}
\includegraphics[width=1.6\textwidth, clip, bb= 0 0 2026 1127]{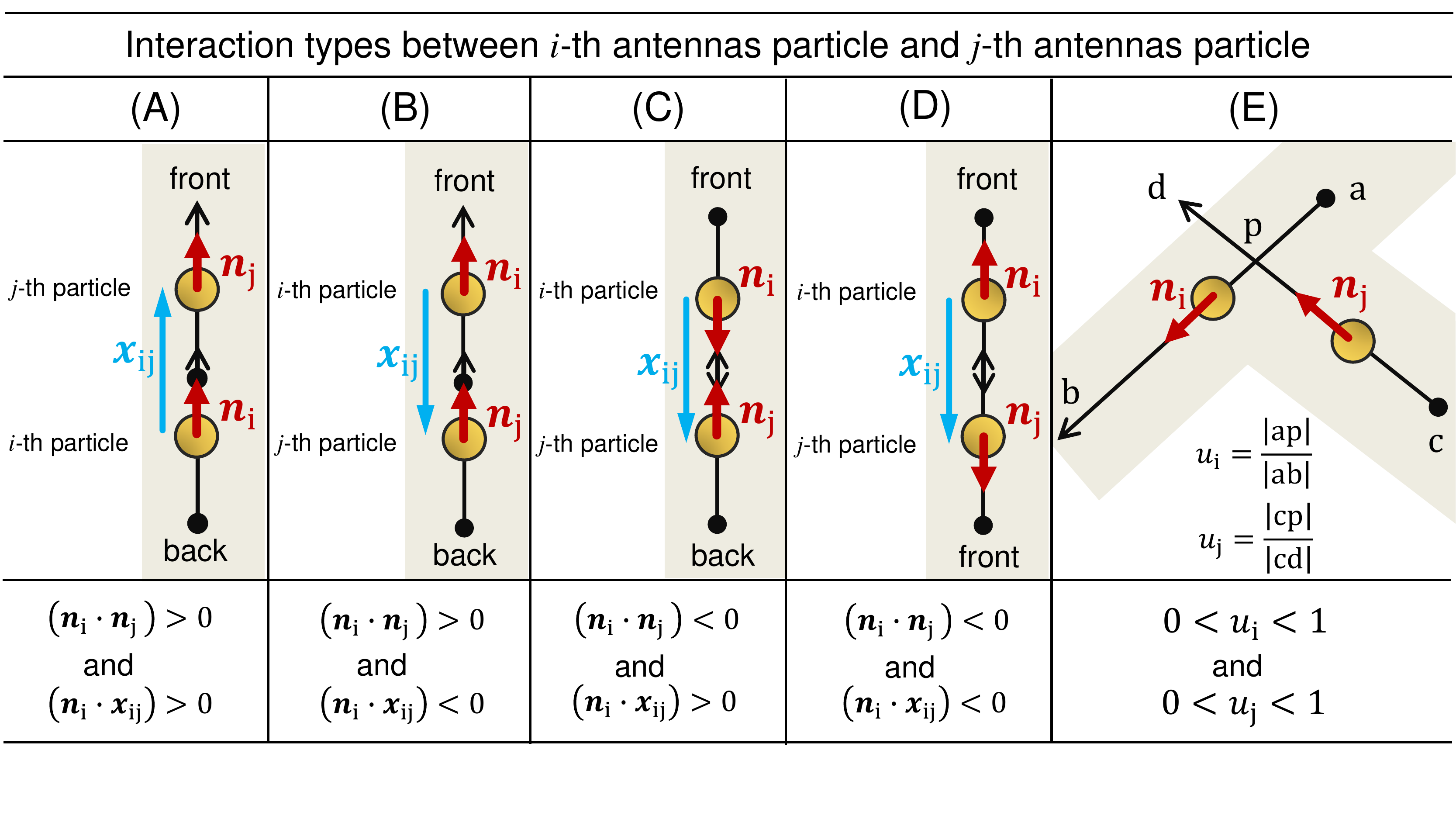}
\caption{Interaction types between $i$-th antenna particle and $j$-th antenna particle on the target system.}
\label{fig:approachype3}
\end{figure*}

In case $\phi$ does not meet the relationship shown in Eq.(\ref{eq:onsameroad}), we decide that 
the $i$-th and $j$-th particles are not on an identical line, which suggests that these two particles are on different crossing paths. 
In such a case, we judge the collisions among these antennas by using the line-line collision detection algorithm used in polyhedral geometrics.
Let the edge positions of the antennas of the $i$-th particle be ${\bf a}({a}_{x}, {a}_{y})$ and ${\bf b}({b}_{x}, {b}_{y})$, and 
let those of the $j$-th particle be ${\bf c}({c}_{x}, {c}_{y})$ and ${\bf d}({d}_{x}, {d}_{y})$. 
At this time, the arbitrary point of ${\bf p}_{\rm ab}$ on the line ${\bf ab}$ 
and that of ${\bf p}_{\rm cd}$ on the line ${\bf cd}$ are expressed as
\begin{eqnarray}
\vec{\bf p}_{\rm ab} &=& {\bf a} + {u}_{\rm i}({\bf b}-{\bf a}) \label{eq:lineeq1}\\
\vec{\bf p}_{\rm cd} &=& {\bf c} + {u}_{\rm j}({\bf d}-{\bf c}) \label{eq:lineeq2}
\end{eqnarray}
Here, ${u}_{\rm i}$ and ${u}_{\rm j}$ indicate the scalar value uniquely determined by each point.
In case line ${\bf ab}$ and line ${\bf cd}$ cross, the left-hand side of Eq.(\ref{eq:lineeq1}) 
and that of Eq.(\ref{eq:lineeq2}) correspond to each other.
At this time, we obtain the unique values of ${u}_{\rm i}$ and ${u}_{\rm j}$ 
by solving the simultaneous linear system of Eq.(\ref{eq:lineeq1}) and Eq.(\ref{eq:lineeq2}).
The cross of two lines only occurs when $0 < {u}_{\rm i} < 1$ and $0 < {u}_{\rm j} < 1$:
\begin{description}
\item [(E)]~If $0 < {u}_{\rm i} < 1$ and $0 < {u}_{\rm j} < 1$, then:\\ 
The $i$-th particle and $j$-th particle collide with each other. 
The system issues pause instructions to the either of the two particles according to the priority rule defined at the joint.
\end{description}

Unlike the case that both particles exist on the same lane, we do not divide the interaction (E) into four subtypes of interactions by their normal vectors because every corner is separated into two paths by the single checkpoint. Namely, once the interaction (E) is detected, the system identifies the exact corner by the location of the particles and pauses the particle that belongs to the rear path of the corner. 
Figure~\ref{fig:approachype3} summarizes equations and the schematics of all interaction types from (A) to (E).

In our model, the system issues multiple instructions to the particles when they go on to the forks. (e.g., the forks around checkpoint C in Fig.\ref{fig:schemview2}). In this case, the particles pause when at least one instruction given to them suggests a pause.
For better understanding, we describe the final decision-making of a particle using the following formula.
Let the total number of particles approaching the $i$-th particle be $N_{\rm i}$; set a serial number to 
the approaching particles between $1$ and $N_{\rm i}$. 
Donate a binary function of the $m$-th particle of the $N_{\rm i}$ particles as $I_{\rm i}^{\rm m}$, which returns one when the system issues pause to the $i$-th particle, and it returns zero for other cases. The final decision-making $D_{\rm i}$ of the $i$-th particle is obtained as
\begin{eqnarray}
{I}_{\rm i}^{\rm s} &=& \sum_{m \in N_{\rm i}} I_{\rm i}^{\rm m} \label{eq:dm1}\\
D_{\rm i} &=& \begin{cases}
~{\rm true}~~~( {I}_{\rm i}^{s}\ne 0) \label{eq:dm2}\\
~{\rm false}~~({\rm otherwise})
\end{cases}
\end{eqnarray}
In case that $D_{\rm i}$ is $\rm true$, the $i$-th particle pauses.

\section{SIMULATIONS}\label{seq:simulation}
\subsection{Definitions of the physical properties}\label{seq:defphyspara}
In this paper, we examine three physical properties: the throughput $Q$ of the whole system, changing rate of schedule $S$, and averaged occupation of the parking sites $U_{\rm T}$ of the target system. First, we define throughput $Q$ as
\begin{eqnarray}
Q &:=& \frac{N_{\rm exit}}{N_{\rm plan}}
\end{eqnarray}
Here, $N_{\rm exit}$ indicates the number of particles that exit from outlet $\rm L$ of Fig.\ref{fig:schemview2}
 during the simulation. $N_{\rm plan}$ shows the number of particles that plan to enter the system from inlet A each day.

Second, we define $N_{\rm chg}$ as the number of particles that changes the destination of the parking site when going through either checkpoint E or F (E for the domestic route and F for the international route). We define the changing rate of schedule $S$ as
\begin{eqnarray}
S &:=& \frac{N_{\rm chg}}{N_{\rm exit}}
\end{eqnarray}

In the end, we describe $U_{\rm D}$ as the number of occupied parking sites in parking lane $G_{\rm D}$ and $U_{\rm I}$ as that of the occupied parking sites in parking lane $G_{\rm I}$. $U_{\rm D}$ and $U_{\rm I}$ are averaged by the total time step $N_{\rm t}$. 
We define $U_{\rm T}$ as
\begin{eqnarray}
U_{\rm T} &:=& U_{\rm D} + U_{\rm I} \label{eq:ocuppiedsites}
\end{eqnarray}

\subsection{Setting of input parameters}\label{seq:inputpara}
We set the length of a square cell on the checkerboard to $2.2$ m and the number of cells of $(L_{\rm x}, L_{\rm y})$ to $(382, 1,483)$ so that our system fits the real geometry of the targetting airport. We adopt a parallel update method; the system updates all cells on the checkerboard simultaneously. We set the physical time of a time step to be $1$~s and set the number of total time steps $N_{\rm t}$ to $86,400$ for each case, which corresponds to $24$~h of a day. We perform simulations for different cases of the distribution parameter $\alpha$ in between $0$ and $1.0$ at intervals of $0.05$. We carry out $15$ cases in each simulation. As mentioned in Section 2.2, we investigate the system by setting the staying time $T_{\rm sp}$ in both cases, with stochastic and with deterministic parameters. In the former case, we set the pair of stochastic parameters $(\bar{\tau}_{\rm s}, \bar{\sigma}_{\rm s})$ of $T_{\rm sp}$ to $(60~{\rm min}, 10~{\rm min})$, aiming to reproduce the typical staying time at real airport terminals. In the latter case, we set $(\bar{\tau}_{\rm s}, \bar{\sigma}_{\rm s})$ to $(60~{\rm min}, 0~{\rm min})$. 
We set the interval velocities by calculating the substep at each time step; 30 substeps for path AB and path KL, 15 substeps for path BC and path GK, and 3 substeps for other paths. Each of these values corresponds to 237.6~km/h, 118.8~km/h, and 23.7~km/h in the physical scale, respectively.
We carry out each simulation for different cases of intervals of arrival time $T_{\rm in}$ in $2$ min, $3$ min, and $4$ min.

\begin{figure}[t]
\vspace{-2.0cm}
\includegraphics[width=0.6\textwidth, clip, bb= 0 0 1280 1280]{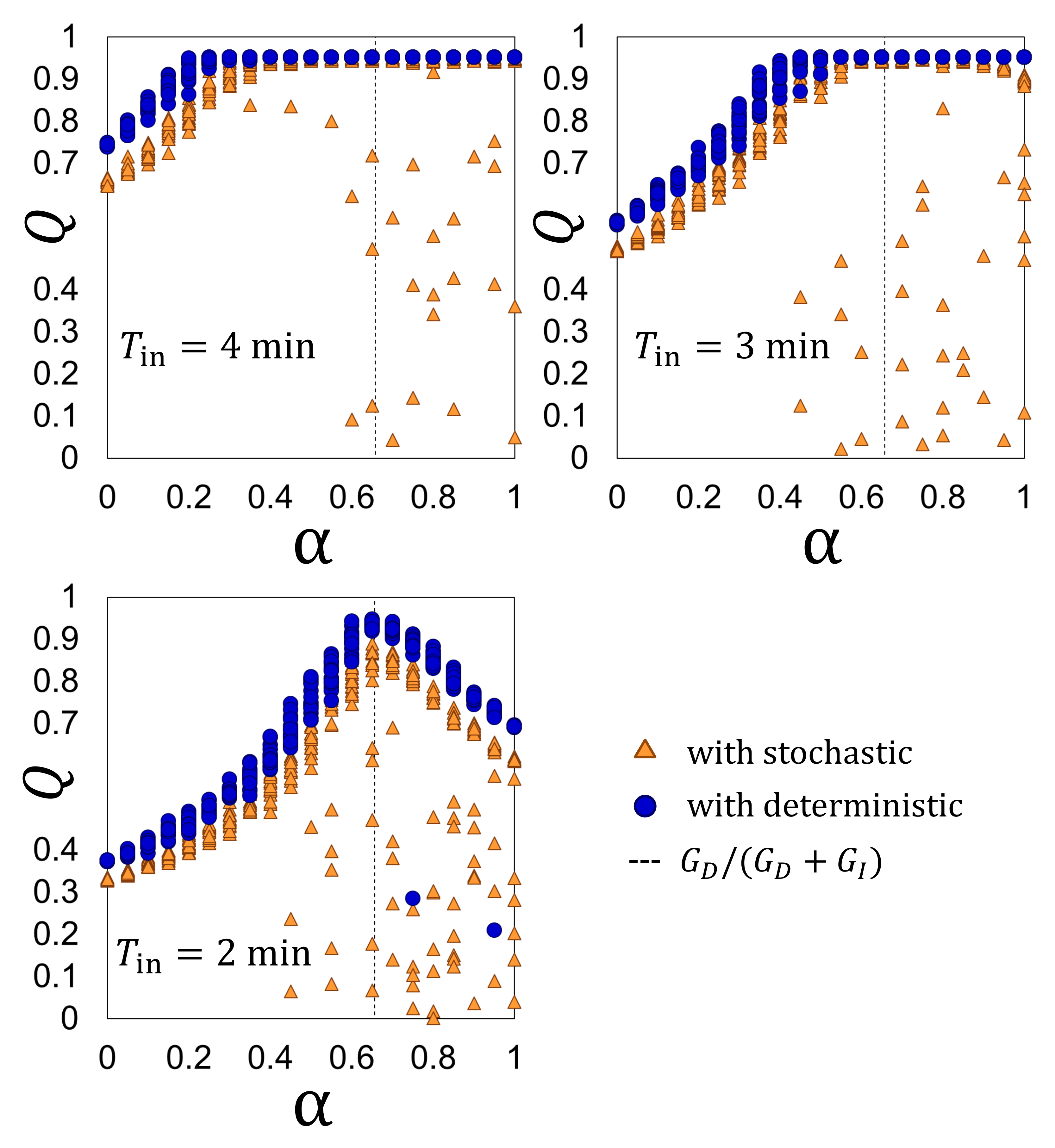}
\caption{Dependence of throughput $Q$ for the whole system on the probability $\alpha$ that distributes the particles to the domestic route $R_{\rm D}$.}
\label{fig:throughput}
\end{figure}
\subsection{Simulation results}\label{seq:simresults}
Figure~\ref{fig:throughput} shows the dependence of throughput $Q$ of the whole system on probability $\alpha$, which distributes the particles to the domestic route $R_{\rm D}$, for different cases of the arrival time interval $T_{\rm in}$~(from $2~{\rm min}$ to $4~{\rm min}$). The blue-colored circle symbol shows the results when setting $T_{\rm sp}$ with deterministic parameters. The orange-colored triangle indicates results in the case of setting $T_{sp}$ with stochastic parameters using the half-normal distribution. The dashed line shows the ratio of the number of parking sites of the domestic route $G_{\rm D}$ to that of the total system ($G_{\rm D}+G_{\rm I}$). This ratio corresponds to the maximum capacity of the parking lane $G_{\rm D}$. In case of setting $T_{\rm in}$ to $4~{\rm min}$, the throughput $Q$ increases as probability $\alpha$ increases, and it reaches a plateau after $\alpha$ becomes more larger than a specific value. We observe similar phenomena in the case of $3~{\rm min}$. 

In the case of $2~{\rm min}$, an important observation was made; the throughput $Q$ reduces after $\alpha$ reaches the maximum capacity. The overall reduction of throughput $Q$ in between $0.6$ and $1.0$ seems to be unique because the throughput plateaus after saturation, in similar systems~\cite{SIMAIAKIS2014251, Khadilkar2013, Eun2016}. Besides, we confirmed from Fig.\ref{fig:throughput} that the existence of the delay of staying time is not the critical factor for the overall reduction because we observe it in both cases: with stochastic and with deterministic parameters. Indeed, the delay of $T_{\rm sp}$ causes a substantial deterioration of throughput $Q$ in exceptional cases; however, it has little effect on the mainstream behavior of the whole system. 

\begin{figure}[t]
\vspace{-2.0cm}
\includegraphics[width=0.6\textwidth, clip, bb= 0 0 1280 1280]{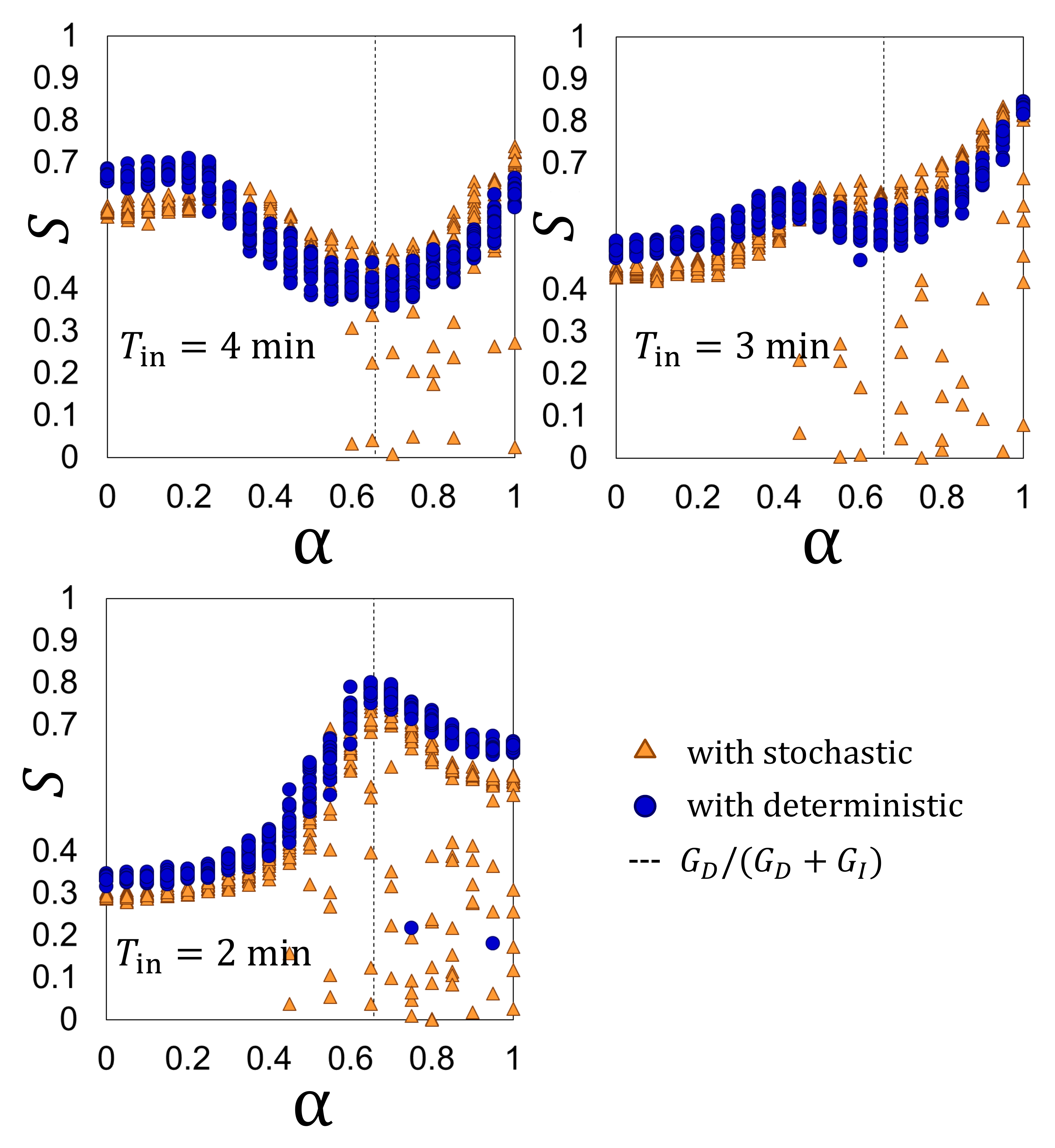}
\caption{Dependence of the changing rate $S$ for the whole system on the probability $\alpha$ that distributes the particles to the domestic route $R_{\rm D}$.}
\label{fig:changingrate}
\end{figure}

Figure~\ref{fig:changingrate} shows the dependence of the changing rate $S$ of the whole system on probability $\alpha$ that distributes the particles to the domestic route $R_{\rm D}$, for different cases of arrival time interval $T_{\rm in}$ from $2~{\rm min}$ to $4~{\rm min}$. Each symbol is the same in Fig.\ref{fig:throughput}. The results show further unexpected behaviors. It is not easy to find the apparent dependence on parameter $\alpha$ in every case. Furthermore, the curves are entirely different among the three cases of setting $T_{\rm in}$ to $2~{\rm min}$, $3~{\rm min}$, and $4~{\rm min}$ regardless of whether it is with stochastic or with deterministic parameters.
Hereafter, we focus on the deterministic cases to clarify the reasons for the phenomena observed in throughput $Q$ and changing rate $S$.

\begin{figure}[t]
\vspace{-2.0cm}
\includegraphics[width=0.6\textwidth, clip, bb= 0 0 1280 1280]{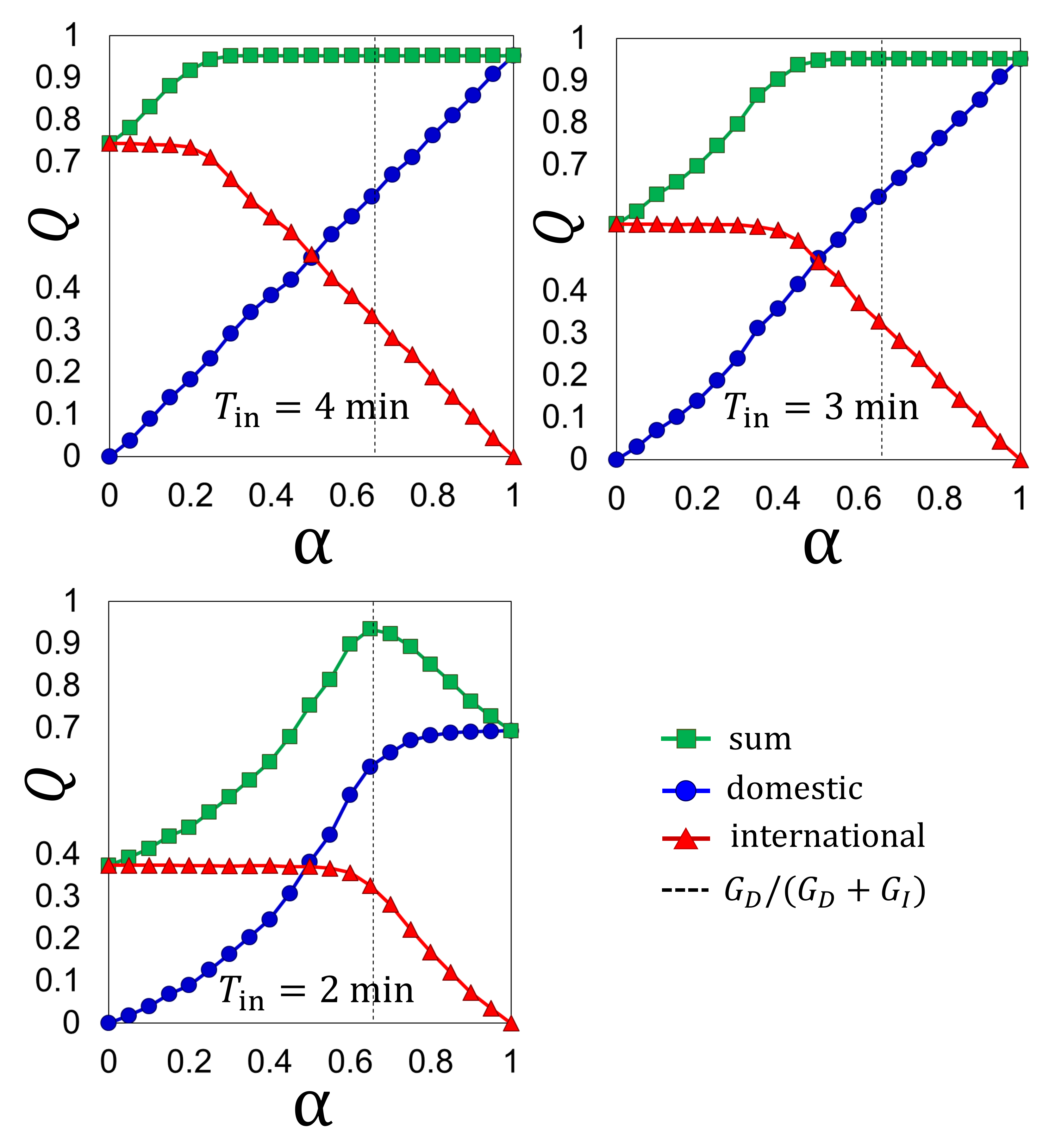}
\caption{Sample mean of the throughput $Q$ in case of setting $T_{\rm sp}$ with deterministic parameters, as shown in Fig.\ref{fig:throughput}, and its breakdown.}
\label{fig:throughput-average}
\end{figure}

An obvious reason was found attributable to these unique behaviors. Figure~\ref{fig:throughput-average} shows the sample mean of the throughput displayed in Fig.\ref{fig:throughput} (the solid line with the green-square symbol) and its breakdown: the throughput of the domestic route $R_{\rm D}$ (the solid line with the blue-colored circle symbol) and that of the international route $R_{\rm I}$ (the solid line with the red-colored triangle symbol).
As shown in Fig.\ref{fig:throughput-average}, the sum of the throughputs of the domestic route $R_{\rm D}$ and the international route $R_{\rm I}$ becomes the throughput of the whole system. In the case of $4~{\rm min}$, the throughput of the international route $R_{\rm I}$ shows an almost plateau when the probability $\alpha < 0.2$ because of the saturation of the parking lane, and it decreases when $\alpha > 0.2$. Meanwhile, the throughput of the domestic route $R_{\rm D}$ increases because the amount of particles increases proportional to probability $\alpha$. The same is almost true in the case of $3~{\rm min}$. These linear increases can be attributed to the fact that the traffic of particles does not surpass the capacity of the domestic lane $G_{\rm D}$ in these cases.
In the case of $2~{\rm min}$, the throughput of the international route $R_{\rm I}$ decreases similarly as that in the cases of $3~{\rm min}$ and $4~{\rm min}$. On the other hand, the throughput of the domestic route $R_{\rm D}$ increases until it reaches the maximum capacity of the domestic lane $G_{\rm D}$, and after that, it saturates unlike the cases of $3~{\rm min}$ and $4~{\rm min}$. 

\begin{figure}[t]
\vspace{-2.0cm}
\includegraphics[width=0.6\textwidth, clip, bb= 0 0 1280 1280]{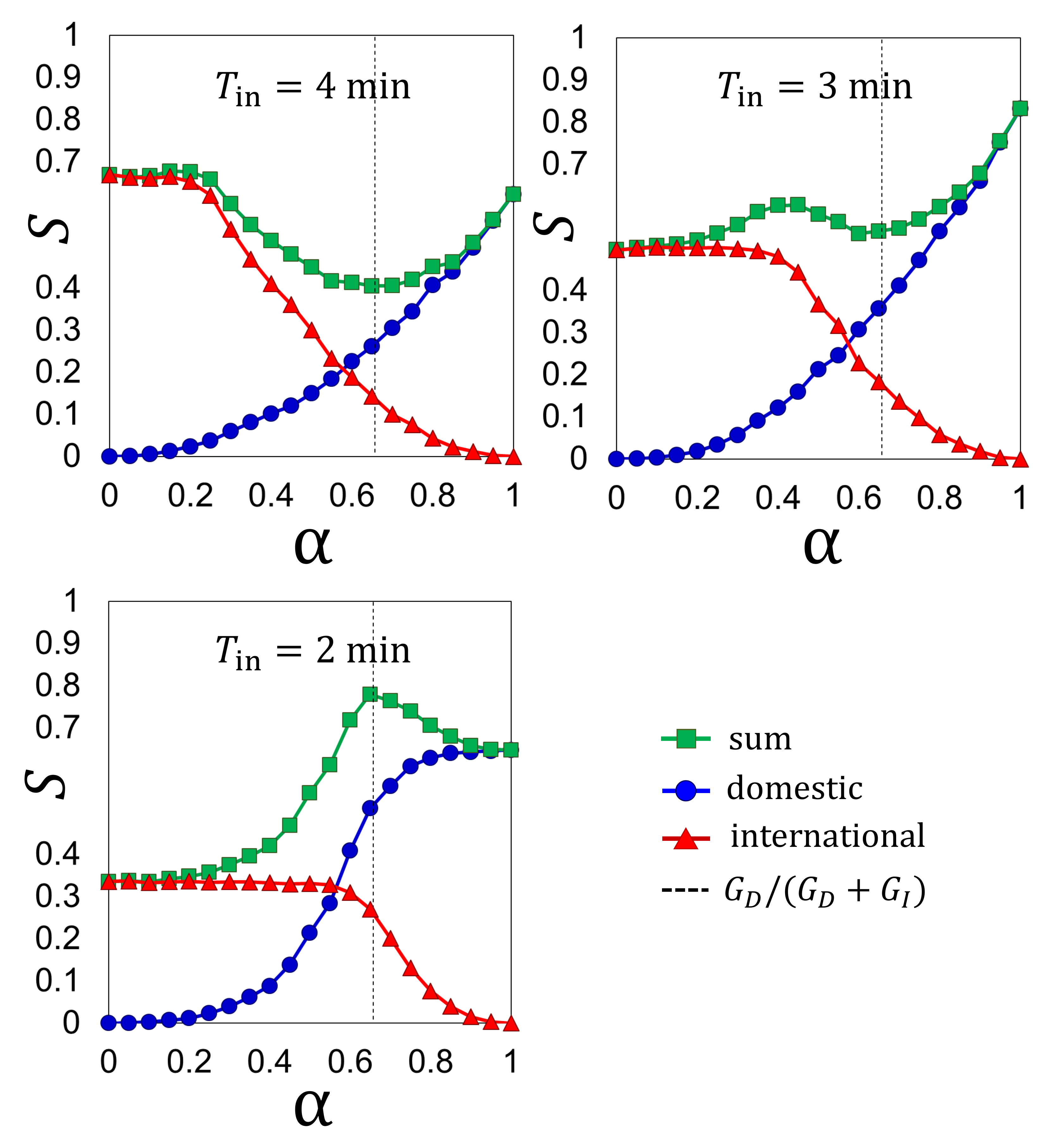}
\caption{Sample mean of the changing rate $S$ in case of setting $T_{\rm sp}$ with deterministic parameters as whown in Fig.\ref{fig:throughput}, and its breakdown.}
\label{fig:changingrate-average}
\end{figure}

\HL{Simply put, the reduction of the throughput of the whole system shown in the case of $2~{\rm min}$ is described as follows. As the distribution ratio of the particles to the international route $R_{\rm I}$ decreases, the throughput of the international route $R_{\rm I}$ reduces while at the same time, that of the domestic route $R_{\rm D}$ saturates, the simultaneous effect of which causes the reduction in the throughput of the whole system.}
A similar explanation can be provided for every case of the changing rate $S$, as shown in Fig.\ref{fig:changingrate-average}. Although each route simply shows either monotonic decrease, monotonic increase, or saturation, their sum displays very complex behavior. 

\subsection{Analysis with the classical $M/M/c$ queue}\label{seq:comparemmc}
The simulation results in Section~\ref{seq:simresults} indicate that the function of the queueing system of each route independently contributes to the whole system. In this section, we compare the simulation results of the average occupied time with the classical queueing model by Erlang~\cite{ERLANG-A-K} (For the details of the classical theoretical queueing model, refer to~\cite{doi:10.1002/0471200581.ch7, Gross:2008:FQT:1972549}). \HL{Since our system has a finite number of service windows (the parking sites) in each route, we apply the $M/M/c$ queueing model~\cite{doi:10.1002/0471200581.ch7} to each route of the target system as the first-order approximation level.}

In the classical $M/M/c$ queue, the average number of occupied sites among $c$ sites corresponds to the ratio of the arrival rate $\lambda$ to the service rate $\mu$. Since the target system directs each particle to the domestic route $R_{\rm D}$ with probability $\alpha$, the effective arrival rates of route $R_{\rm D}$ and route $R_{\rm I}$ become $\alpha\lambda$ and $(1-\alpha) \lambda$, respectively. 

In case of $c=1$, the arrival rate $\lambda$ and the service rate $\mu$ respectively correspond to the inversed values of the interval of arrival time $T_{in}$ and the staying time $T_{\rm sp}$ ($\lambda = T_{\rm in}^{-1}$, $\mu=T_{sp}^{-1}$) as long as we ignore the effect of the congestion. In this approximation, we regard $T_{\rm in}^{-1}$ and $T_{\rm sp}^{-1}$ as the arrival rate $\lambda$ and service rate $\mu$ as for the $M/M/1$ queue. Because the number of occupied sites does not suppress the maximum capacity, $U_{\rm D}$ and $U_{\rm I}$ in each route can be expressed by the classical $M/M/c$ queueing model, as follows:
\begin{eqnarray}
U_{\rm D} &=& {\rm min}\Biggl[ N_{\rm D}, \frac{\alpha T_{\rm sp}}{T_{\rm in}}\Biggr] \label{eq:mmc:1}\\
U_{\rm I} &=& {\rm min}\Biggl[ N_{\rm I}, \frac{(1-\alpha) T_{\rm sp}}{T_{\rm in}}\Biggr ] \label{eq:mmc:2}
\end{eqnarray}
Here, $N_{\rm D}$and $N_{\rm I}$ indicate the number of parking sites in the domestic lane $G_{\rm D}$ and that in the international lane $G_{\rm I}$, respectively.
From Eq.(\ref{eq:ocuppiedsites}), Eq.(\ref{eq:mmc:1}), and Eq.(\ref{eq:mmc:2}), we obtain the average number of occupied sites $U_{\rm T}$ as
\begin{eqnarray}
U_{\rm T} &=& {\rm min}\Biggl[ N_{\rm D}, \frac{\alpha T_{\rm sp}}{T_{\rm in}}\Biggr] + {\rm min}\Biggl[ N_{\rm I}, \frac{(1-\alpha) T_{\rm sp}}{T_{\rm in}}\Biggr ] 
\end{eqnarray}

\begin{figure}[t]
\vspace{-2.0cm}
\includegraphics[width=0.6\textwidth, clip, bb= 0 0 1280 1280]{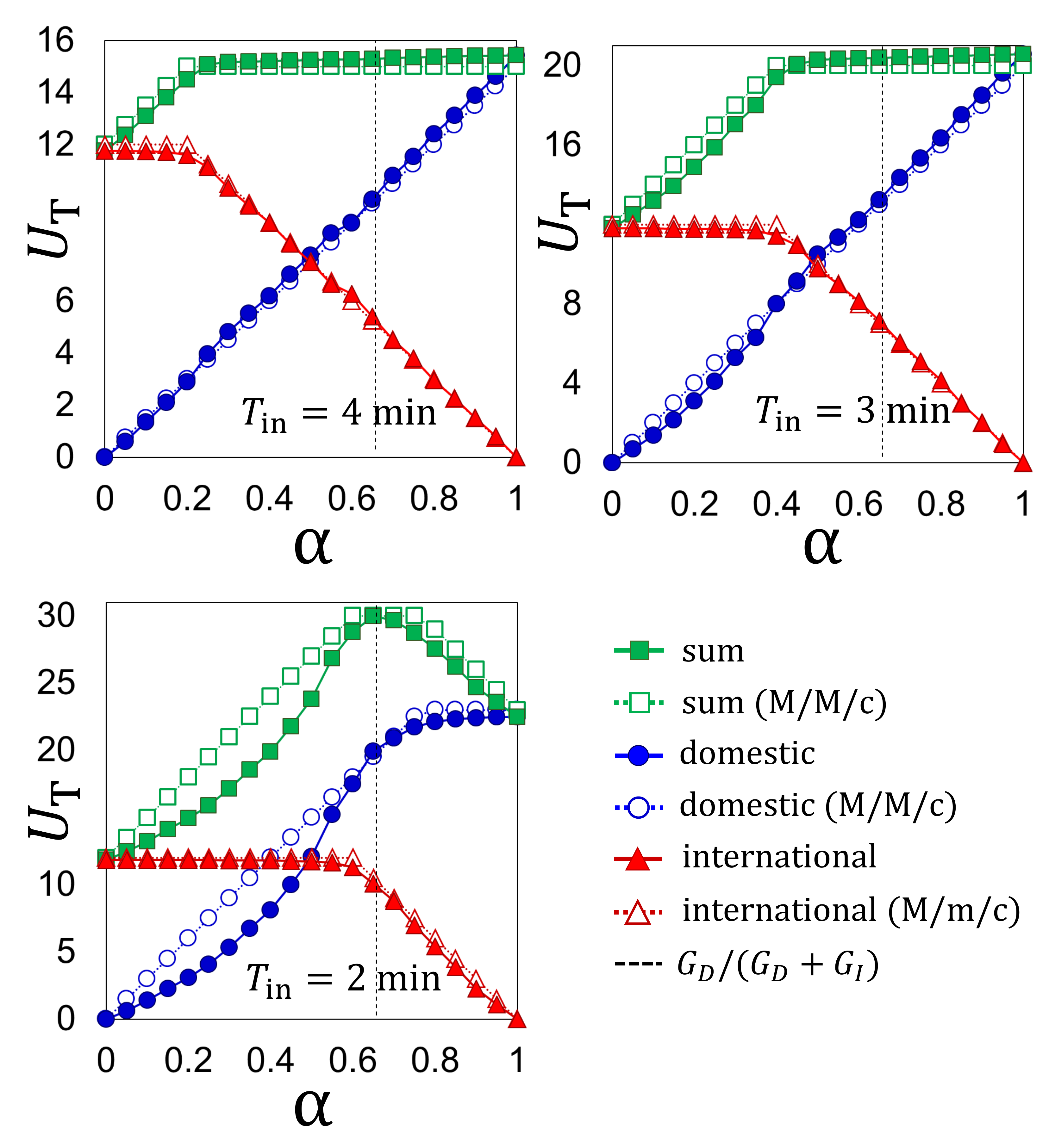}
\caption{Comparison of the average number of occupied sites $U$ with the classical $M/M/c$ queueing theories.}
\label{fig:comparisonwithmmc}
\end{figure}

Figure~\ref{fig:comparisonwithmmc} shows the comparison of the simulations with the approximations by the classical $M/M/c$ queueing models for different cases of the arrival time interval $T_{\rm in}$ between $2~{\rm min}$ and $4~{\rm min}$. The solid lines with the filled symbol indicate the simulations. The dashed lines with the white-colored symbols show the approximations obtained using the $M/M/c$ queueing model. It was found that the $M/M/c$ queueing models describe the simulation results appropriately, and they strongly support the fact that \HL{the simultaneous effect of the decrease in the throughput of the international route $R_{\rm I}$ and the saturation of that of the domestic route $R_{\rm D}$ causes the reduction in the throughput of the whole system in the case of $2~{\rm min}$.}

For further investigations, we describe the following observations. Whereas the simulation results of route $R_{\rm I}$ show good agreement with approximations in all cases from $2$ to $4~{\rm min}$, the simulations of the domestic route $R_{\rm D}$ was observed to deviate from the approximations as $T_{\rm in}$ decreases because of the longer distance of route $R_{\rm D}$ compared to route $R_{\rm I}$, and the lack of the congestion effect in the classical $M/M/c$ queueing model.

\subsection{Discussion for future works}
\begin{figure}[t]
\vspace{-2.0cm}
\includegraphics[width=0.6\textwidth, clip, bb= 0 0 1280 1280]{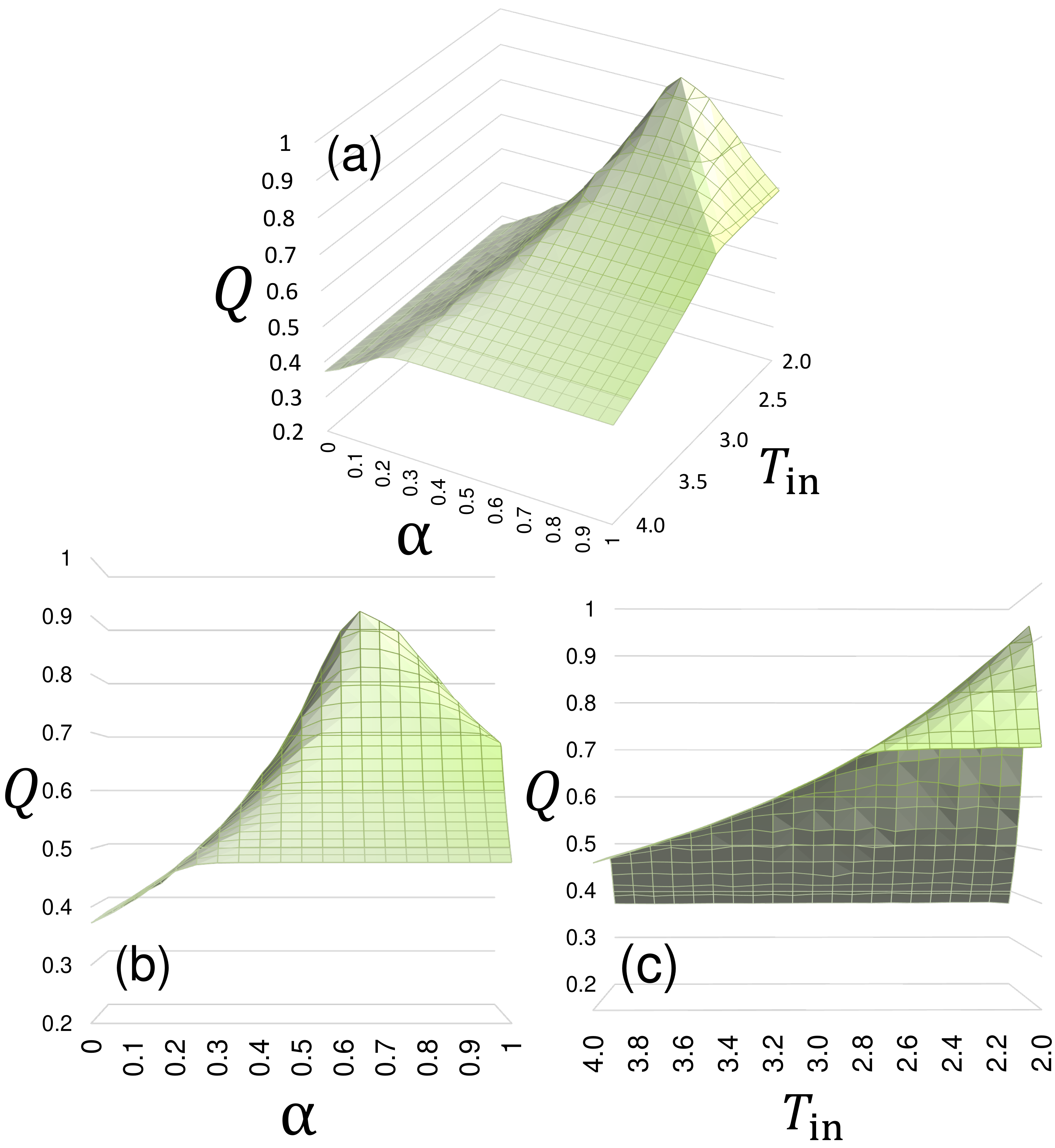}
\caption{3D contour plot of the dependencies of the throughput $Q$ on the distribution probability $\rm alpha$ and the interval of arrival time $T_{\rm in}$ in three types of views: (a) bird's-eye, (b) $\alpha$-side, and (c) $T_{\rm in}$-side.}
\label{fig:contourplot}
\end{figure}

\HL{Figure~\ref{fig:contourplot} shows the three-dimensional contour plot of the dependencies of the throughput $Q$ on the distribution probability $\rm \alpha$ and the interval of arrival time $T_{\rm in}$. It was observed that throughput $Q$ gets sluggish and reaches the plateau as $T_{\rm in}$ increases, thus making a type of sigmoid curve, the tendency of which is similar to the experimental data of the same type of systems in previously reported works~\cite{SIMAIAKIS2014251, Khadilkar2013, Eun2016}. On the other hand, the dependence of throughput $Q$ on arrival time interval $T_{\rm in}$ does not reach the plateau in between $4$ and $2~{\rm min}$ at around where the parameter $\alpha$ gets close to the ratio of the number of parking sites in the domestic route to that in the whole system. It should be emphasized that, the $\alpha$-dependence of the physical properties of this kind of system was first discussed in this paper. Particularly, we found that the simultaneous effect of the decrease, increase, and saturation of the physical properties in the two different routes causes unusual behaviors of the whole system. Since the targeting system fixes the number of parking sites in this study, there might exist a possibility to find another interesting fact caused by the aforementioned simultaneous effects in different situations; these other facts should be studied in future works.} 
 
\section{CONCLUSIONS}
We introduced a two-dimensional stochastic lattice model that comprises a junction of two traveling lanes, each of which has a parking lane. In our model, each particle has a pair of antennas in the moving direction to detect other approaching particles. 
As a case study, we abstracted a target system from a real-world airport that has one of each domestic and international terminals.
We applied our model to the target system and investigated the physical characteristics thoroughly. 
The contribution of this paper is as follows:

The proposed system was observed to display interesting behavior. The throughput shows a sudden decrease after reaching the maximum capacity of the parking lane. On the other hand, the dependence of the changing rate of schedule on the system parameter disorderly fluctuates, and it hides its dependence on the parameter. Our simulations and approximations by the M/M/c queueing model clearly explain these phenomena and support the following fact. \HL{As the distribution ratio of particles to the international route decreases, the throughput of the international route reduces; simultaneously, the throughput of the domestic route $R_{\rm D}$ saturates. This simultaneous effect of the decrease and saturation causes reduction in the throughput of the whole system.}

\section*{Acknowledgements}
This research was supported by MEXT as ``Post-K Computer Exploratory Challenges'' (Exploratory Challenge 2: Construction of Models for Interaction Among Multiple Socioeconomic Phenomena, Model Development and its Applications for Enabling Robust and Optimized Social Transportation Systems)(Project ID: hp190163), partly supported by JSPS KAKENHI Grant Numbers 25287026, 15K17583 and 19K21528.
\\\\
\appendix
\section{Visualizations}
\begin{figure*}[t]
\vspace{-5.0cm}
\hspace{+0.3cm}
\includegraphics[width=1.3\textwidth, clip, bb= 0 0 1280 720]{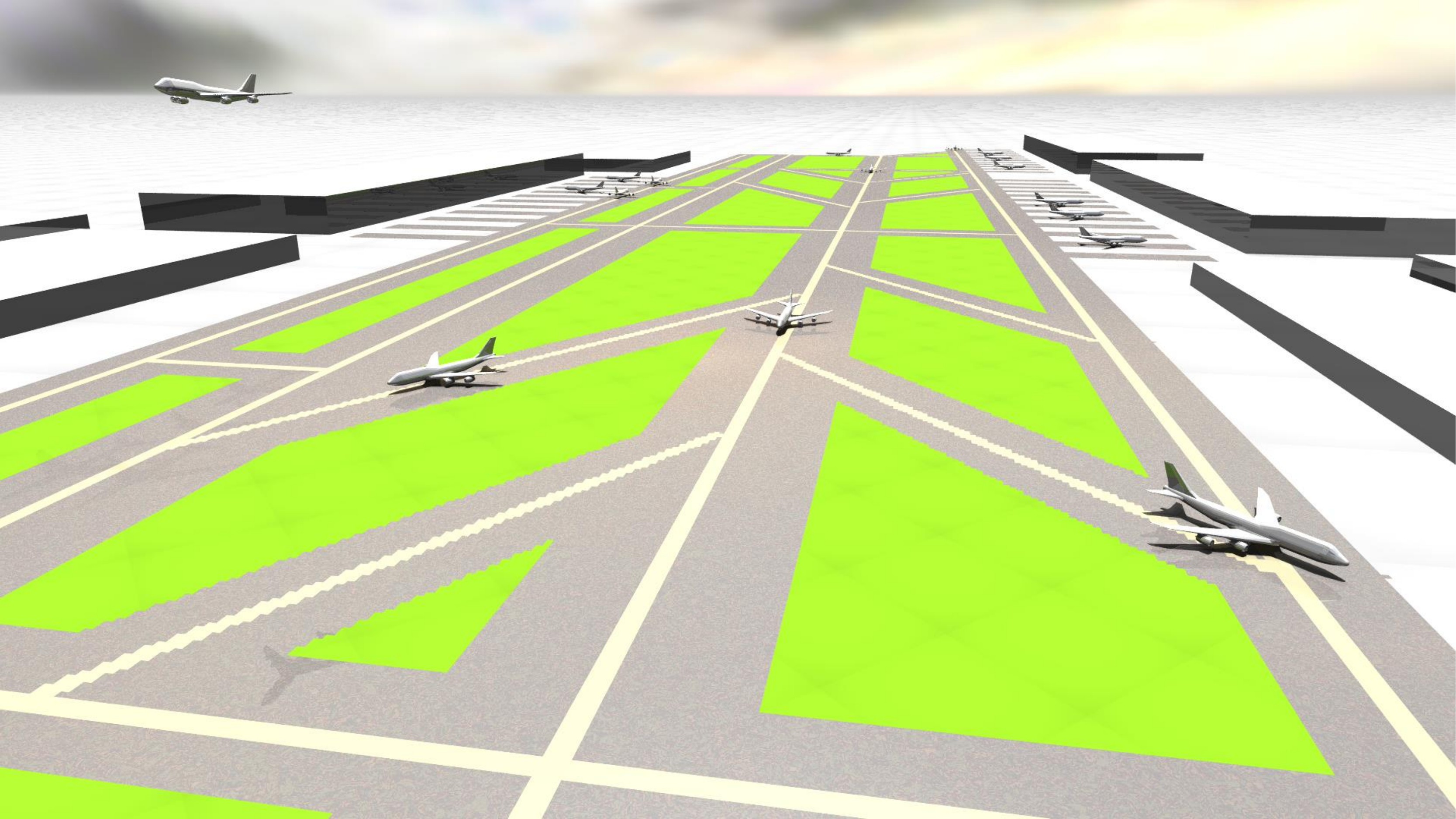}
\caption{The realistic visualization of an air-ground traffic simulation in Section~\ref{seq:simulation}.}
\label{fig:realisticvis}
\end{figure*}

Figure~\ref{fig:realisticvis} exhibits the realistic visualization of an air-ground traffic simulation in Section~\ref{seq:simulation}.
\vspace{-1.0cm}
\bibliographystyle{h-physrev3}
\bibliography{reference}

\clearpage




\end{document}